# Analytical Approach to Study the Impacts of Mutual Coupling on Transmission Lines Protection Systems

Azade Brahman [(1)], Dilan Novosad [(1)], Mehriar Tabrizi [(1)], Tim Cook [(2)], Wei Jen Lee, Fellow IEEE [(3)]
(1) DNV GL – Energy Advisory, Dallas Texas
(2) Cross Texas Transmission, Austin Texas
(3) Department of Electrical Engineering, University of Texas at Arlington
azade.brahman@dnvgl.com

*Abstract*— While there are numerous literatures that have addressed the impact of mutual coupling on the reliability and security of protection schemes and have provided possible mitigation solutions, there has not been adequate research and documentation presenting a comprehensive analytical approach to 1) estimate the magnitude of mutual coupling and 2) quantify the adverse impact of mutual coupling in real-life scenarios under several system faults across various types of protective elements. This should be considered as the first stage of any mutual coupling related study preceding the second stage in which the mitigation against mutual coupling is to be developed. The proposed methodology can be used to study the impact of mutual coupling on ground overcurrent relays, ground and phase distance as well as pilot protection schemes.  As part of the proposed approach, EMT simulation is utilized to quantify the extent of sub-transient overshoot and current reversal that may have adverse impact on the performance of studied relays.  A real-life case study within the ERCOT network has been used to demonstrate the proposed study approach.

## I. Introduction

In power transmission networks, there could be locations in which two or more transmission circuits are sharing right of way or common tower structures along the transmission path. This "parallel" configuration can be seen in various structural scenarios including, but not limited to, double-circuit lines on the same tower structure or single-circuit lines running in parallel in narrow corridors or even on common tower due to financial or spatial constraints. [1],[2].

In energized parallel configurations, each parallel circuit, in addition to the self-generated magnetic flux, experiences the alternating magnetic flux generated by the other parallel circuit. This is called "mutual coupling" and leads to induction of zero-sequence currents and voltages on both parallel circuits. The strength of mutual coupling has an inverse relation with the spacing between the parallel circuits and direct relation with the length of the parallel sections. The induced zero-sequence current, if not accounted for, may cause challenges for protection, control and operation personnel [1], [3], [4].

Since mutual coupling effect results in induction of zero-sequence current and voltages, it is expected that it only affects the protection schemes against ground faults, by either altering the measured fault current magnitude or direction [1]. Thus, compensation methods are required in setting the ground directional overcurrent, distance and directional comparison elements of a protection system, associated with parallel transmission lines [1]-[4].

While number of solutions have been proposed to avoid mis-operation of protective elements in presence of mutual coupling not all of these solutions are robust enough to mitigate the adverse impact of mutual coupling under all the practical scenarios[1]-[4].

A tutorial on protection schemes and recommendations on relay setting and compensation methods in presence of mutual coupling effect is provided in [1], [3]. References [2] and [5] study the application of negative-sequence component as the polarizing quantity in ground directional elements in the presence of mutual coupling. This is warranted primarily due to the negligible magnitude of negative-sequence mutual coupling impedance.

While authors in [6] have mainly focused on double-circuit lines and how associated protective elements can be improved to mitigate against mutual coupling effect, [4] addresses other possible configurations that result in mutual coupling and its impact on ground distance, ground overcurrent and directional protection schemes. Furthermore, the calculation of current flowing through the transmission line under various operating and topological conditions considering the effects of mutual coupling event has been provided in [7].

In addition to improvements and recommendations around ground directional overcurrent and distance relays, there are literatures that have considered use of other form of protection schemes to address the adverse impacts of mutual coupling. A current differential protection scheme is developed in [8] using transmission line π-equivalent model and phase coordination approach. Adaptive digital distance relaying scheme has been proposed by [9],[10] to mitigate against the drawbacks associated with conventional ground distance elements in the presence of mutual coupling. It has been argued that the

This research was supported by DNV GL Energy Advisory

proposed digital distance relaying schemes can measure the correct magnitude of apparent impedance in presence of mutual impedance and fault resistance during an inter-circuit LL and LLG faults.

The mutual coupling can affect the power system reliability and resiliency through mis-operation of relays on critical transmission lines. The coherence among the generators are highly depended to the network structure [11]-[13] and inappropriate trip caused by mutual coupling can have negative impact in dynamic stability. Moreover, although some of the special protection and mitigation counter measures [14]-[15] are not directly affected by mutual coupling phenomena but, the performance of these unit might be affected through false trips of transmission lines and consequent changes in the impedance characteristic of the system. While the system reliability and resilience improvement are usually evaluated against deliberate intrusions [16] and extreme events [17], or pursed through substation structure enhancement [18], [19], mitigating the mutual coupling can improve network reliance. In addition, the system operation and safety which are traditionally improved by optimization approach [20], [21], protective device coordination and arc flash analysis [22] can also be affected by mutual coupling.

While there are numerous literatures that have addressed the impact of mutual coupling on the reliability and security of protection schemes and have provided possible mitigation solutions, there has not been adequate research and documentation presenting a comprehensive analytical approach to 1) estimate the magnitude of mutual coupling and 2) quantify the adverse impact of mutual coupling in real-life scenarios under several system faults across various types of protective elements. This should be considered as the first stage of any mutual coupling related study preceding the second stage in which the mitigation against mutual coupling is to be developed.

This paper presents an analytical approach to study the impact of mutual coupling induced zero-sequence components on the operation of protection systems. A case study has been developed within the ERCOT network to evaluate the performance of directional Ground Overcurrent (GOC), Ground and Phase Distance elements as well as pilot protection schemes for neighboring transmission lines running in close proximity of one another.

## II. THEORY AND BACKGROUND

Ideally, the impedance matrix consists of only positive and zero-sequence impedances. However, in reality, some extra terms caused by the coupling effect of parallel lines is also reflected in impedance matrix. Zero-sequence coupling effect can provide a mutual impedance up to 50-70% of the self-impedance, while the effect of positive and negative-sequence component on the adjacent line is negligible [2].

Self and mutual impedances of transmission lines are discussed in detail in [1], [23]. Carson's equation is used to derive the mathematical equivalent for self and mutual impedances. The self-impedance of a conductor with returning ground path and the mutual impedance of any two or more parallel conductors running in close vicinity with common returning earth path are given as follows:

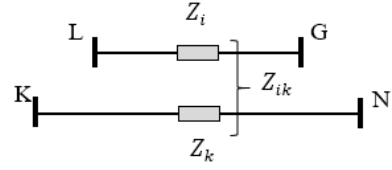

Fig. 1. Two Parallel Lines with Mutual and Self Impedances

$$Z_{ik} = r_{ik} + j0.0012562f \left[ \ln\left(\frac{1}{D_{ik}}\right) \right] ohms/Km \quad (1)$$

Here $f$ represents frequency and:
- For $i = k$:
  $r_{ik} = Conductor's\ ac\ resistance$
  $D_{ik} = GMR\ of\ conductor\ i$
- For $i \neq k$: $r_{ik} = 0$
  $D_{ik} = Distance\ between\ conductors\ i\ \&\ k$

As depicted in Fig. 2, considering the voltage drop equations for a three-phase system and substituting the Carson's equation, self and mutual impedances of a three-phase system can be derived as demonstrated in (2) [2].

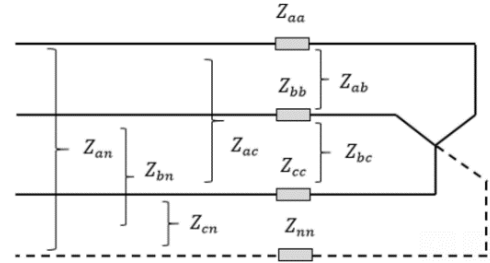

Fig. 2. Mutual Impedances Among Three Phase Transmission Lines

$$Z_{ik} = r_{ik} + 0.0009865f \\ + j0.0012562f \left[ \ln\left(\frac{1}{D_{ik}}\right) + 6.7458 \right] ohms/Km \quad (2)$$

The self and mutual impedance as obtained in (2) can be used to derive the system impedance matrix as illustrated in (3) to be utilized for protection and fault calculation purposes.

$$Z_{abc} = \begin{bmatrix} Z_{aa} & Z_{ab} & Z_{ca} \\ Z_{ab} & Z_{bb} & Z_{bc} \\ Z_{ca} & Z_{bc} & Z_{cc} \end{bmatrix} \quad (3)$$

Subsequently, the transmission line symmetrical components of the self and mutual impedances are then derived using the formula as shown in (4).

$$Z_{seq} = A^{-1} Z_{abc} A \quad (4)$$

In effect, mutual impedances induce a voltage on each mutually coupled line equal to the product of the mutual impedance and the current flow in the parallel line. In various articles, this is explained as an induced zero sequence current in parallel lines which run in close proximity. Due to the presence of a mutual impedance, the induced current is analogous to the current induced in the secondary of a transformer. To this extent, a general demonstration of two parallel lines in Fig. 1 can be represented using an ideal transformer as an illustration for mutual impedances as depicted in Fig. 3 [1]-[3].

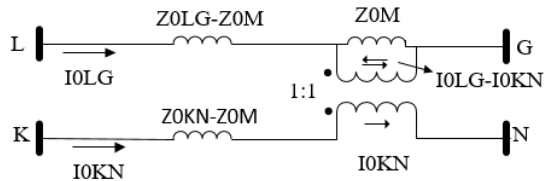

Fig. 3. Equivalent Network for Mutual Coupling Effect

The voltage drop across each parallel line in both examples are the same and equal to:

$$V_{LG} = Z_{0LG}I_{0LG} + Z_{0M}I_{0KN} \quad (5)$$

$$V_{KN} = Z_{0KN}I_{0KN} + Z_{0M}I_{0LG} \quad (6)$$

The amount of mutual impedance between the two coupled lines is inversely proportional to the spacing between the two lines while is directly in proportion to the length of the coupled section. Therefore, by increasing the length of coupled section and decreasing the proximity of the two lines, a considerable increase in the mutual impedance $Z_{0M}$ is derived.

## III. CASE STUDY

### A. Model Development

#### 1) Transmission Line Modeling

Case study within the ERCOT network was developed to assess the impact of mutual coupling between various existing transmission lines and a future planned double circuit transmission line. The model development effort requires two key steps. The first step involves transmission line model development using a frequency dependent model in PSCAD/EMTDC software to quantify the mutual coupling impedances. The key information used in model development are listed as follows:

- Distance between phase conductors
- Distance between phase and ground conductors
- Distance between the earth surface and phase/ground conductors
- Radius and DC resistance of phase and ground conductors
- Phase conductor bundle specifications and Earth resistivity

Another critical parameter in estimation of mutual coupling impedances is the horizontal separation distance between each of the neighboring transmission lines included for analysis. The separation distances have been incorporated into the PSCAD/EMTDC models. The line constant computation module in PSCAD/EMTDC is then utilized to estimate the mutual coupling impedances between the proposed line and each of the existing lines.

#### 2) Short Circuit Model

The second step is to develop an ERCOT short-circuit model in ASPEN Oneliner to conduct the mutual coupling study. The mutual coupling impedances calculated as part of the first step were incrementally modeled in the study case. Additionally, protective relay elements as well as pilot protection schemes for neighboring facilities were included in the study model. The final short-circuit case is utilized to assess the impact of mutual coupling on the operation of protective relay elements and pilot protection scheme.

### B. Impact on Directional GOC Relays

Directional ground OC relay elements in the case study operate on residual current of $3I_0$ i.e. three times of zero-sequence current measured by relay. The $3I_0$ operating quantity in conjunction with presence of AC connection between the coupled lines introduces certain complications while conducting the mutual coupling study. The operating current $3I_0$ as measured by nearby relays comprises of two (2) components:

- Fault based $3I_0$: When an unbalanced fault involving ground occurs on the line, current of $3I_0$ will circulate through the neighboring transmission lines given the presence of electrical connection between these transmission facilities. Magnitude of this current is solely a function of electrical connection between the neighboring lines. This current would exist even if there were no mutual coupling.
- Induced $3I_0$: Presence of mutual coupling will induce an additional $3I_0$ component on the neighboring transmission lines during fault conditions.

The induced $3I_0$ component that occurs due to mutual coupling may be additive or subtractive in polarity to fault based $3I_0$. Consequently, the total operating current $3I_0$ that is measured by ground OC relays may increase or decrease due to presence of mutual coupling.

The Assessment of Directional Ground Overcurrent (GOC) relays was performed considering two different fault scenarios:

- Placing the faults along the nearby double circuit transmission line
- Placing line-end faults at each ends of the nearby double circuit transmission line

In each of these two scenarios, the operating time of the relays on the coupled sections as well as the fault current seen by the relays have been captured before and after the inclusion of mutual coupling. Fig. 4 serves to illustrate the difference between fault-based and induced components of the operating current $3I_0$. It is evident that the operating current $3I_0$ is significantly increased due to presence of mutual coupling. In this scenario, there is a possibility that the pick-up time of the ground OC relay will be significantly reduced.

Table I depicts results of this analysis for ground OC relays for the above illustration. In case of most study scenarios,

presence of mutual coupling induces relatively higher $3I_0$ current as measured causing faster relay trip time. Further, there are specific instances where the relay pick-up time in presence of mutual coupling is slower in comparison to the case with no mutual coupling. In such situations, the mutually induced $3I_0$ was observed to be in opposite polarity to the fault based $3I_0$.

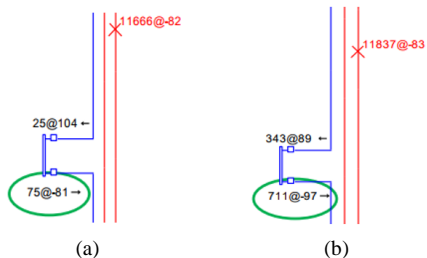

Fig. 4. $3I_0$ observed (a) Before mutual coupling (b) After mutual coupling

TABLE I
GOC RELAY OPERATIONS WITH AND WITHOUT MUTUAL COUPLING

| Fault Location Along New Double Ckt | Primary Protection | | Backup Protection | |
|---|---|---|---|---|
| | Without MC Trip Time (s) | With MC Trip Time (s) | Without MC Trip Time (s) | With MC Trip Time (s) |
| 10% | 9999 | 4.1 | 9999 | 5.3 |
| 20% | 9999 | 2.6 | 9999 | 3.4 |
| 30% | 9999 | 1.9 | 9999 | 2.5 |
| 40% | 9999 | 2.0 | 9999 | 2.7 |
| 50% | 9999 | 3.5 | 9999 | 4.6 |
| 60% | 9999 | 11.7 | 9999 | 15.3 |
| 70% | 9999 | 9999 | 9999 | 9999 |
| 80% | 9999 | 38.9 | 9999 | 50.9 |
| 90% | 22.7 | 9999 | 29.7 | 9999 |

Instances may arise in which line-end faults on the new double circuit line are not instantaneously cleared by remote-end due to protection system failures. Therefore, the remote-end relay may clear the fault at a delayed time based on Zone 2 setting. Table II depicts the GOC operating time for situations as discussed.

The results associated with the assessment indicate that GOC relays on neighboring transmission lines are affected due to zero-sequence mutual coupling causing the pick-up times to be significantly reduced or increased depending on the fault location.

TABLE II
GOC RELAY OPERATIONS FOR LINE-END FAULTS WITH AND WITHOUT MUTUAL COUPLING

| Fault Location Along New Double Ckt | Primary Protection | | Backup Protection | |
|---|---|---|---|---|
| | Without MC Trip Time (s) | With MC Trip Time (s) | Without MC Trip Time (s) | With MC Trip Time (s) |
| Near End Opened | | | | |
| 0.5% | 9999 | 14.7 | 9999 | 19.2 |
| 5% | 9999 | 11.0 | 9999 | 14.3 |
| 10% | 9999 | 8.3 | 9999 | 10.8 |
| 15% | 9999 | 6.4 | 9999 | 8.4 |
| 20% | 9999 | 5.1 | 9999 | 6.6 |
| Remote End Opened | | | | |
| 80% | 9999 | 9999 | 9999 | 9999 |
| 85% | 9999 | 9999 | 9999 | 9999 |
| 90% | 9999 | 9999 | 9999 | 9999 |
| 95% | 9999 | 9999 | 9999 | 9999 |
| 99.5% | 9999 | 9999 | 9999 | 9999 |

*C. Impact on Ground and Phase Distance Relays*

Ground and Phase distance relays are also included as part of the case study to identify any potential mis-operation in presence of mutual coupling. Ground distance relays operate based on the zero-sequence current component of a fault. By introducing mutual coupling, ground distance relays may trip for faults that are beyond the zone reach point (over-reach) or restrain from tripping for faults that are within the reach point (under-reach). In a similar manner, phase distance relays are included as part of the assessment to identify potential mis-operation of the relays vis-à-vis over- and/or under- reach pickup conditions.

Specific fault conditions were studied as part of the assessment:

- Single Line to Ground (SLG) for Ground Distance Relays
- Double Line to Ground (LLG) for Phase Distance Relays

It is important to note, line-line and/or 3-phase faults do not lead to a zero-sequence quantity within the network and therefore has not been included in the analysis.

SLG and LLG faults were placed along the protected transmission element to identify the Zone 1 reach points for ground and phase distance relays, respectively, in absence of mutual coupling. The procedure was repeated in presence of mutual coupling to assess any potential over- and/or under-reach for each of the relays. Two study conditions associated with the status of the mutually coupled transmission line were included as part of the analysis:

- Mutually coupled line in-service
- Mutually coupled line out-of-service and grounded at both ends

The results associated with the ground and phase distance relay assessment indicate that the Zone 1 reach-point is not observed to change for each of the study conditions. The case study concludes that potential for over- and/or under- reach pickup conditions due to the inclusion of mutual coupling is not expected to occur for the ground and phase distance relays under study.

*D. Impact on Pilot Protection Scheme*

Pilot protection used for lines provides the possibilities of high-speed simultaneous detection of phase- and ground-fault protection for 100% of protected section from all terminals [2]. As a part of differential protection, the Pilot protection element can be used in all voltage levels within the network and uses a communication channel to compare the terminal quantity and send an appropriate signal to maintain stability by clearing the fault in the shortest possible time. In practice, pilot schemes are used to ensure that under internal fault conditions the protective relay elements operate simultaneously to clear the fault, however mis-operation of the same would be destructive to the system operation.

Failure to recognize the location of the fault may cause the pilot scheme to mis-operate. This is prevalent among mutually coupled lines as the induced zero-sequence components may alter the current and impedance magnitude of the line. In presence of mutual coupling, pilot protection schemes such as

DCB and POTT may falsely interpret the direction of the fault and cause instantaneous tripping of nearby relays.

As part of this effort, the operation of DCB and POTT was monitored in presence of mutual coupling under various fault locations. Based on the results of the analysis, the operation of the above-mentioned pilot protection schemes was not observed to be impacted for relays which are polarized using negative-sequence components.

*E. EMTDC Result Validation*

For each of the study scenarios outlined above, the study results represent a steady state (i.e. post transient) solution and does not capture any subtransient response that may occur during faulted conditions. A steady state solution is not capable of capturing any DC offset or current reversal which may occur during the subtransient interval after the fault. A significant level of DC offset may cause the relay trip time to reduce as the relay will observe a higher fault current. Similarly, a reversal in current during the fault results in a subtransient change in polarity and may cause the mis-operation of protection systems. Pilot protection schemes such as POTT and DCB schemes may falsely interpret the fault to be in the forward direction which could potentially cause instantaneous tripping. Based on this, EMT simulations were conducted in PSCAD/EMTDC software to capture the subtransient response during faulted conditions. The subtransient response observed as part of this effort is then used to validate the previous obtained study results.

The study model was developed using an equivalent simplified model to be simulated in PSCAD/EMTDC software. EMT simulations were performed under Single Line to Ground (SLG) fault based conditions for various locations in vicinity of the mutually coupled transmission lines. Several fault locations were studied to ensure all corresponding relays are properly polarized and correctly identify the direction of the fault.

For illustrative purposes, Fig. 5 depicts the current response as measured by a relay protecting a neighboring transmission facility. As seen in Fig. 5, the overshoot observed immediately after occurrence of the fault is not significantly larger that the post-transient component. The amount of overshoot is limited given the system strength of the study region. The study region is strongly interconnected and yields high fault current levels promoting the strength of the system. Based on this, the subtransient current does not demonstrate a significant amount of overshoot and quickly dampens out to a steady state solution.

In addition, the PSCAD model was utilized to determine any relay mis-operation caused by current reversal leading to an incorrect relay polarization. Each of the ground OC relays included as part of this analysis are polarized using negative sequence components. Each relay uses the following polarization method to determine the direction of the fault:

- Forward Fault Detection: Current Leads the Voltage
- Reverse Fault Detection: Current Lags the Voltage

Fig. 5 depicts that the negative sequence current leads the negative sequence voltage. Therefore, the relay under study identifies the fault location in the reverse direction. The abovementioned procedure was repeated for various fault scenarios to identify whether mutual coupling was observed to cause any current reversal and mis-operation with respect to the polarization of the relay.

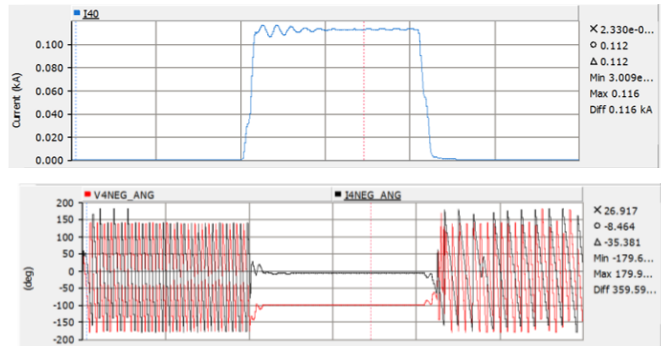

Fig. 5. Zero-Sequence Current for Relay Under Study

None of the EMT simulation results were indicative of any current reversal (polarity change) or significant sub-transient overshoot to the extent that it impacts the performance of the above-mentioned relays.

IV. CONCLUSIONS

This paper presents a comprehensive analytical approach that can be used by system planners to evaluate and quantify the impact of mutual coupling on the operation of protection systems under various system faults. A real-life case study within the ERCOT network has been used to demonstrate the proposed study approach. The proposed methodology consists of detailed modeling of mutual coupling as well as analytical approach to study the impact of mutual coupling on GOC relays, ground and phase distance as well as pilot protection schemes. Lastly, as part of the proposed approach, EMT simulation has been utilized to quantify the extent of sub-transient over shoot and current reversal that may have adverse impact on the performance of such relays.